\documentclass[11pt]{article}
\usepackage{vietnam,epsfig}

\def\be{\begin{equation}}
\def\ee{\end{equation}}
\def\bea{\begin{eqnarray}}
\def\eea{\end{eqnarray}}

\begin{document}
\vspace*{4cm}
\title{PHOTO-ASSISTED SHOT NOISE IN COULOMB INTERACTING SYSTEMS}

\author{A. CR\'EPIEUX, M. GUIGOU, A. POPOFF, P. DEVILLARD, AND T. MARTIN}

\address{Centre de Physique Th\'eorique, CNRS Luminy case 907, 13288 Marseille cedex 9, France}

\maketitle

\abstracts{
We consider the fluctuations of the electrical current (shot noise) in the presence of a voltage time-modulation. For a non-interacting metal, it is known that the derivative of the photo-assisted noise has a staircase behavior. In the presence of Coulomb interactions, we show that the photo-assisted noise presents a more complex profile, in particular for the two following systems: 1) a two-dimensional electron gas in the fractional quantum Hall regime for which we have obtained evenly spaced singularities in the noise derivative, with a spacing related to the filling factor and, 2) a carbon nanotube for which a smoothed staircase in the noise derivative is obtained.}

%%%%%%%%%%%%%%%%%%%%%%%%%%%%%%%%%%%%%%%%%%%%%%%%%%%%%%%%%%%%%%%%%%%%%%%%%%%%%%%%%%
%
%
%	INTRODUCTION
%
%
%%%%%%%%%%%%%%%%%%%%%%%%%%%%%%%%%%%%%%%%%%%%%%%%%%%%%%%%%%%%%%%%%%%%%%%%%%%%%%%%%%%

\section{Introduction}

In mesoscopic systems, the measurement of 
shot noise makes it possible to probe the effective charges
which flow in conductors. 
This has been illustrated experimentally and theoretically when the interaction between electrons is 
less important \cite{reznikov} or when it is more relevant \cite{kane_fisher_noise,saminadayar}. Additional informations can be obtained through the photo-assisted noise when an AC bias is superposed to the DC bias.
Experimentally, photo-assisted noise has been measured in diffusive wires, diffusive junctions and quantum point contacts \cite{schoelkopf}. For normal metals, the noise 
derivative displays steps~\cite{lesovik} at integer values of the ratio $\omega_0/\omega$,
where $\omega$ is the AC frequency and $\omega_0$ is related to the DC voltage. We naturally expect that this behavior
is modified in Coulomb interacting systems. The present work deals with two specific one-dimensional correlated systems:~a Hall bar 
in the fractional quantum Hall regime, for which charge transport occurs 
via two counter-propagating chiral edges states, and a carbon nanotube to which electrons are injected from a Scanning Tunneling Microscope~(STM) tip. 

%%%%%%%%%%%%%%%%%%%%%%%%%%%%%%%%%%%%%%%%%%%%%%%%%%%%%%%%%%%%%%%%%%%%%%%%%%%%%%%%%%
%
%
%	EHQF
%
%
%%%%%%%%%%%%%%%%%%%%%%%%%%%%%%%%%%%%%%%%%%%%%%%%%%%%%%%%%%%%%%%%%%%%%%%%%%%%%%%%%%%

\section{Photo-assisted noise in the fractional quantum Hall regime}

The first system we consider is a two-dimensional electrons gas in the fractional quantum Hall regime which is described by the Hamiltonian $H=H_0+H_B$. The kinetic term $H_0=(\hbar \mathrm{v}_F/4\pi)\sum_{r=R,L}\int ds \partial_{s}\phi_r(t))^2$ describes the right and left moving chiral excitations along the edge states ($\phi_{R}$ and $\phi_L$ are the bosonic fields), and
 $H_B=A(t)\Psi_R^\dag(t)\Psi_L(t)+h.c.$ describes the transfer of quasiparticles from one edge to the other.
$\Psi_{R(L)}=F_{R(L)}e^{i\sqrt{\nu}\phi_{R(L)}(t)}/\sqrt{2\pi a}$ where $F_{R(L)}$ is a Klein factor, $a$, the short-distance cutoff and $\nu$, the filling factor which characterizes the charge $e^*=\nu e$ of the backscattered quasiparticles. The backscattering amplitude between the edge states has a time dependence due to the applied voltage $V(t)=V_0+V_1\cos(\omega t)$:
\begin{eqnarray}\label{A}
A(t)=\Gamma_0 e^{i\frac{\nu e}{\hbar}\int dt V(t)}=\Gamma_0e^{i\omega_0 t+i\frac{\omega_1}{\omega}\sin(\omega t)}=\Gamma_0\sum_{p=-\infty}^{+\infty}J_p\left(\frac{\omega_1}{\omega}\right)e^{i(\omega_0+p\omega)t}~,
\end{eqnarray}

where $\omega_0\equiv \nu e V_0/\hbar$ and $\omega_1\equiv \nu e V_1/\hbar$. We have made an expansion in term of an infinite sum of Bessel functions $J_p$ of order $p$, which is a signature of photo-assisted processes \cite{platero}.

The backscattering current noise correlator is expressed with the help of the Keldysh contour:
$S(t,t')=\sum_{\eta=\pm 1}\langle T_K\{I_B(t^\eta)I_B(t'^{-\eta})\exp(-i\int_K dt_1H_B(t_1))\}\rangle/2$
where $I_B(t)=i\nu eA(t)\Psi_R^\dagger(t)\Psi_L(t)-h.c.$ is the current operator. We are interested in the Poissonian limit only, so 
in the weak backscattering case, one collects the second order contribution in 
the tunnel amplitude $A(t)$.

The main purpose of this work is to analyze the double Fourier transform of the noise $S(\Omega_1,\Omega_2)\propto\int dt\int dt' S(t,t')\exp(i\Omega_1t+i\Omega_2t')$ when both 
frequencies $\Omega_1$ and $\Omega_2$ are set to zero. At zero temperature, the shot noise exhibits divergences \cite{crepieux} at each integer value of the ratio $\omega_0/\omega$ which are not physical since they appear in a range of frequencies where the perturbative calculation turns out to be no more valid. At finite temperature, the photo-assisted noise reads:
\begin{eqnarray}
S(0,0)
&=&\frac{(e^*)^2\Gamma_0^2}{2\pi^2a^2{\overline{\Gamma}}(2\nu)}
\left(\frac{a}{\mathrm{v}_F}\right)^{2\nu}
\left(\frac{2\pi}{\beta}\right)^{2\nu-1}\nonumber\\
&&\times\sum_{p=-\infty}^{+\infty}J_p^2\left(\frac{ \omega_1}{\omega}\right)
\cosh\left(\frac{(\omega_0+p\omega)\beta}{2}\right)\left|{\overline{ \Gamma}}\left(\nu+i\frac{(\omega_0+p\omega)\beta}{2\pi}\right)\right|^2~,
\end{eqnarray}

where $\overline{\Gamma}$ is the Gamma function and $\beta=1/k_BT$.

\begin{figure}[h]
  \includegraphics[height=.22\textheight]{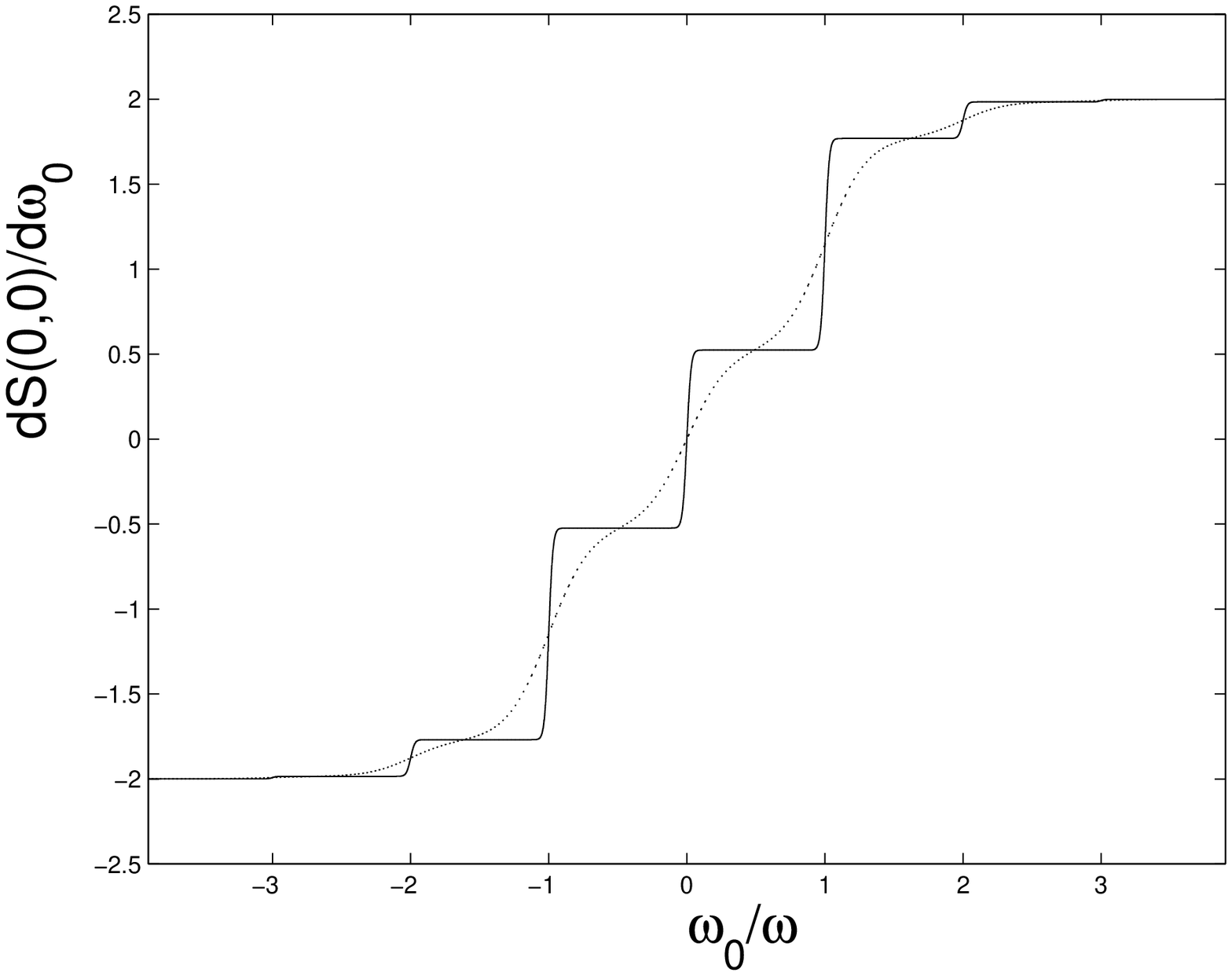}
  \hspace*{1cm}
    \includegraphics[height=.22\textheight]{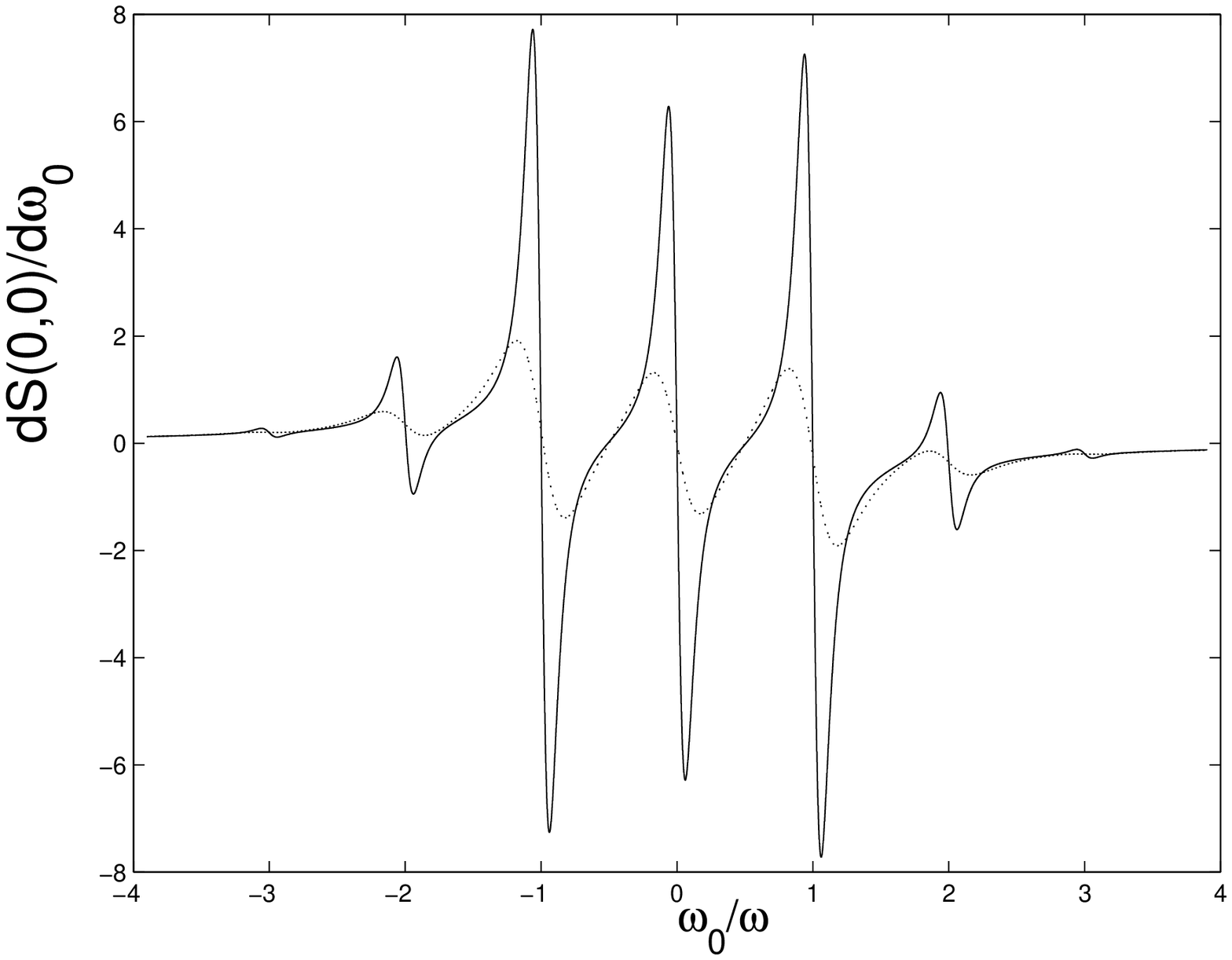}
  \caption{Noise derivative in the fractional quantum Hall regime as a function of $\omega_0/\omega=\nu V_0/\hbar\omega$ for (left) $\nu=1$ at temperatures: $k_BT/\hbar\omega=0.01$ (solid line) and $k_BT/\hbar\omega=0.1$ (dashed line) and for (right) $\nu=1/3$ at temperatures: $k_BT/\hbar\omega=0.05$ (solid line) and $k_BT/\hbar\omega=0.15$ (dashed line). We take $\omega_1/\omega=eV_1/\hbar\omega=3/2$.}
\end{figure}

We have tested the validity of our result by setting $\nu=1$ which corresponds to a non-interacting system. The derivative of the noise according to the bias voltage exhibits staircase behavior as shown on Fig.~1 (left). Steps occur every time  $\omega_0$ is an integer multiple of the AC frequency. This is in complete agreement with the results obtained be Lesovik and Levitov for a Fermi liquid~\cite{lesovik}. For a non-integer value of the filling factor ($\nu=1/3$), the shot noise derivative exhibits evenly spaced singularities as seen on Fig.~1 (right), which are reminiscent of the tunneling density of states singularities for Laughlin quasiparticles. The spacing is determined by the quasiparticle charge $\nu e$ and the ratio of the bias voltage with respect to the AC frequency. Photo-assisted transport can thus be considered as a probe for effective charges at such filling factors, and could be used in the study of more complicated fractions of the quantum Hall effect.

%%%%%%%%%%%%%%%%%%%%%%%%%%%%%%%%%%%%%%%%%%%%%%%%%%%%%%%%%%%%%%%%%%%%%%%%%%%%%%%%%%
%
%
%	NANOTUBE
%
%
%%%%%%%%%%%%%%%%%%%%%%%%%%%%%%%%%%%%%%%%%%%%%%%%%%%%%%%%%%%%%%%%%%%%%%%%%%%%%%%%%%%

\section{Photo-assisted noise in carbon nanotube}

\begin{figure}[h]
\begin{center}
\epsfxsize 6.5cm
\epsffile{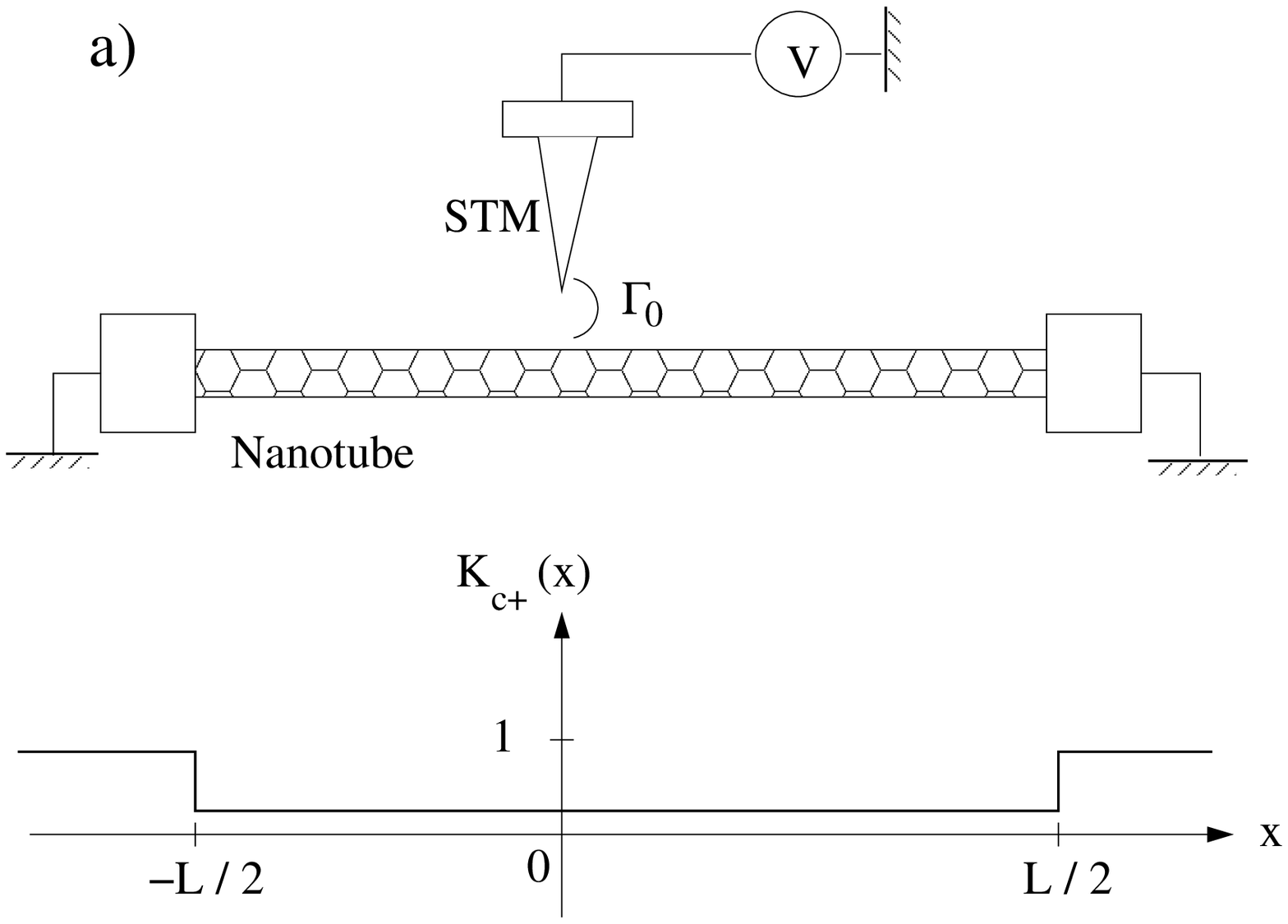}
\hspace*{1cm}
\epsfxsize 7cm
\epsffile{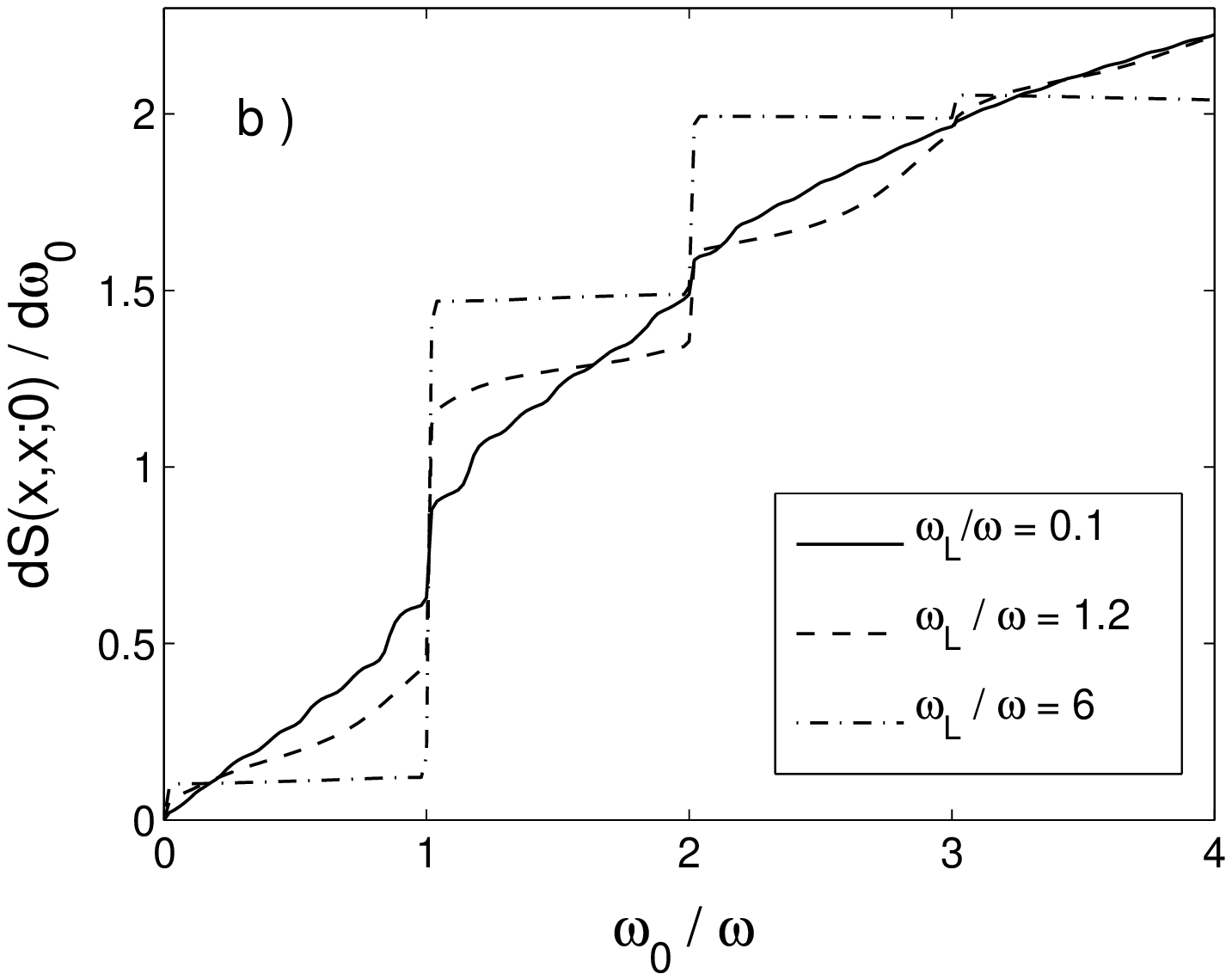}
\caption{a) Picture of the system and spatial variation of the Coulomb interactions parameter $K_{c+}$; b)~Shot noise derivative in the nanotube for $K_{c+}=0.2$, $\omega_1/\omega=2$, $\omega_c/\omega=100$ and several values of $\omega_L/\omega$.\label{fig2}}
\end{center}
\end{figure}

We consider the following setup: an STM tip close to a carbon nanotube connected to leads at both extremities. A voltage applied between the STM and the nanotube allows electrons to tunnel in the center region of the nanotube. As a result, charge excitations propagate along the nanotube towards the right and left leads. This system is described by the Hamiltonian $H=H_\mathrm{N}+H_\mathrm{STM}+H_\mathrm{T}$. The nanotube is a non-chiral Luttinger liquid \cite{egger_98}:

\begin{eqnarray}
      H_\mathrm{N}&=&\frac12\sum\limits_{j\delta}\int\limits_{-\infty}^{+\infty}
      dx\;v_{j\delta}(x)\Bigl[K_{j\delta}(x)(\partial_x\phi_{j\delta})^2
      +K_{j\delta}^{-1}(x)(\partial_x\theta_{j\delta})^2\Bigr]~,
      \label{b_ham}
\end{eqnarray}
where $x$ is the position along the nanotube, $\phi_{j\delta}$ and $\theta_{j\delta}$ are non-chiral bosonic fields and $K_{j\delta}$ is the Coulomb interactions parameter for each charge/spin, total/relative sectors $j\delta\in\{c+,c-,s+,s-\}$. We take $K_{c-}=K_{s+}=K_{s-}=1$, and we assume that $K_{c+}$ depends on position \cite{safi_maslov} as depicted on Fig.~\ref{fig2}a. The velocities satisfy $v_{j\delta}(x)=v_\mathrm{F}/K_{j\delta}(x)$.

The electrons in the metallic STM tip are assumed to be non-interacting. For convenience, the electron field $c_\sigma$ in the STM tip can be described in terms of a semi-infinite Luttinger liquid with Coulomb interactions parameters all equal to one. The tunnel Hamiltonian between the STM tip and the nanotube at position $x=0$ is
$      H_\mathrm{T}(t)=\sum_{r\alpha\sigma}
      \Gamma(t)\Psi_{r\alpha\sigma}^\dagger(0,t)
      c_\sigma(t)+h.c.$
where $r$ corresponds to the branch index, $\alpha$ to the mode index and $\sigma$ to the spin. The fermionic fields for electrons in the nanotube and in 
the STM tip are respectively defined by:
$\Psi_{r\alpha\sigma}(x,t)=F_{r\alpha\sigma}
      \,e^{ik_{\rm\scriptscriptstyle F}rx+
      iq_{\rm\scriptscriptstyle F}\alpha x+i\varphi_{r\alpha\sigma}
      (x,t)}/\sqrt{2\pi a}$ and 
      $c_\sigma(t)=f_\sigma\,e^{i\tilde
      \varphi_\sigma(t)}/\sqrt{2\pi a}$
where $a$ is the ultraviolet cutoff of the Luttinger liquid
model, $F_{r\alpha\sigma}$ and $f_\sigma$ are Klein factors,
$k_{\scriptscriptstyle\rm F}$ is the Fermi momentum and
$q_{\scriptscriptstyle\rm F}$ is the momentum mismatch associated
with the two modes $\alpha$.

In the presence of a voltage modulation superimposed on the constant DC voltage, $V(t)=V_0+V_1\cos(\omega t)$, the tunnel amplitude becomes
$\Gamma(t)=\Gamma\sum_{p=-\infty}^{+\infty}J_p\left(\omega_1/\omega\right)\exp(i(\omega_0+p\omega)t)$,
where $J_p$ the Bessel function on order $p$, $\omega_0\equiv eV_0/\hbar$ and $\omega_1\equiv eV_1/\hbar$. 
The calculation of noise is thus analogous to the one which applies to 
the fractional quantum Hall effect, except that here 
only electrons tunnel in the nanotube.
We thus obtain \cite{guigou}:
\begin{eqnarray}
S(0,0)=\frac{4\Gamma^2 e^2}{\pi^2 a^2} \sum_{p=-\infty}^{+\infty}J^2_p\left(\frac{\omega_1}{\omega}\right)\int_{0}^{+\infty}d\tau\frac{\cos((\omega_0+p\omega)\tau)}{\left(1+\left(\frac{v_F\tau}{a}\right)^2\right)^{\frac{1+\nu}{2}}}\nonumber\\
\times \frac{\cos\left((1+\nu)\arctan\left(\frac{v_F\tau}{a}\right)+\frac{1}{8}\sum_{n=1}^{\infty}\left(\frac{b_{c+}^n}{K_{c+}}+(-b_{c+})^n K_{c+}\right)\arctan\left(\frac{2av_F \tau}{a^2+(nLK_{c+})^2-(v_F\tau)^2}\right)\right)}{\prod_{n=1}^{\infty}\left(\left(\frac{a^2+(nLK_{c+})^2-(v_F \tau)^2}{a^2+(nLK_{c+})^2}\right)^2+\left(\frac{2av_F \tau}{a^2+(nLK_{c+})^2}\right)^2\right)^{\frac{1}{16}\left(\frac{b_{c+}^n}{K_{c+}}+(-b_{c+})^n K_{c+}\right)}}~,
\end{eqnarray}

where $b_{c+}=(K_{c+}-1)/(K_{c+}+1)$ is the reflexion coefficient at the nanotube contacts $x=\pm L/2$ and $\nu=\sum_{j\delta}(1/K_{j\delta}+K_{j\delta})/8$.

The ``standard'' way \cite{lesovik} to display the results for 
photo-assisted transport is to consider the noise derivative as a function 
of voltage: in particular, this allows to compare the results with the 
non-interacting case where the noise derivative exhibits a staircase variation. In Fig.~\ref{fig2}b, we plot the numerically computed noise derivative as a function of the ratio $\omega_0/\omega$ in the presence of Coulomb interactions ($K_{c+}=0.2$).

For $\omega_L/\omega=0.1$ (full line), we are in the limit where the wave packet spatial extension is smaller that the nanotube length. In this regime (except for a small region close to the origin) the vast majority of the voltage scale lies in the regime where $\omega_0>\omega_L$. 
The noise derivative differs from the single electron 
behavior, in the sense that the sharp steps and plateaus expected in this case are absent. 
Instead, because of Coulomb interactions effects, the noise derivative is smoothed out, but there is a clear reminiscence
of the step positions: the slope of $dS(x,x;\Omega=0)/d\omega_0$ increases
abruptly at the location of the steps. We attribute the smoothing to the 
tunneling density of states on the nanotube which is modified by the Coulomb interactions in the nanotube. For $\omega_L/\omega=1.2$ (dashed line), we are in an intermediate regime for which electron wave packets are comparable to the nanotube length. For $\omega_L/\omega=6$ (dashed-dotted line), we are in the limit where electron wave packets are larger than the nanotube length, and as a consequence, the finite size effects dominate over the Coulomb interactions effects and a stepwise behavior in $dS(x,x;\Omega=0)/d\omega_0$, which is typical of non-interacting metals, can be identified. 

%%%%%%%%%%%%%%%%%%%%%%%%%%%%%%%%%%%%%%%%%%%%%%%%%%%%%%%%%%%%%%%%%%%%%%%%%%%%%%%%%%
%
%
%	CONCLUSION
%
%
%%%%%%%%%%%%%%%%%%%%%%%%%%%%%%%%%%%%%%%%%%%%%%%%%%%%%%%%%%%%%%%%%%%%%%%%%%%%%%%%%%%

\section{Conclusion}

Photo-assisted noise is affected by Coulomb interactions in one-dimensional systems. In the fractional quantum Hall effect, the photo-assisted noise shows evenly spaced singularities with a spacing related to the filling factor. As a consequence, photo-assisted noise measurement in such a system could be used to extract fractional charge. In carbon nanotube, Coulomb interactions affect the height and shape of the steps in the differential noise when $\omega_L/\omega$ is small. On the contrary, when $\omega_L/\omega\ge 1$, finite size effects play an important role and attenuate the Coulomb interactions effects.

%%%%%%%%%%%%%%%%%%%%%%%%%%%%%%%%%%%%%%%%%%%%%%%%%%%%%%%%%%%%%%%%%%%%%%%%%%%%%%%%%%
%
%
%	REFERENCES
%
%
%%%%%%%%%%%%%%%%%%%%%%%%%%%%%%%%%%%%%%%%%%%%%%%%%%%%%%%%%%%%%%%%%%%%%%%%%%%%%%%%%%

\end{document}